\documentclass[twocolumn,showpacs,preprintnumbers,amsmath,amssymb,amsfonts]{revtex4}
\usepackage{dcolumn}
\usepackage{bm}
\usepackage{graphicx}

\begin{document}


\title{A new approach to the absorbing boundary conditions for the Schr\"odinger type equations}

\author{M.~Yu.~Trofimov}

\email{trofimov@poi.dvo.ru}
\affiliation{V.~I.~Il'ichev Pacific oceanological institute,\\
 Baltiyskaya St. 43, Vladivostok, 890041, Russia
}%

\
\date{\today}

\begin{abstract}
By the multiple-scale method some new approximate absorbing boundary conditions for the  Schr\"odinger type  equations
are obtained.
\end{abstract}

\maketitle

\section{Introduction}
We will consider the construction of the absorbing boundary conditions for the Schr\"odinger type  equations
(subscripts are used for the derivative with respect to the corresponding variable)
\begin{equation} \label{t1}
\mathrm{i} u_x  + \beta(x) u_{yy} + \nu(x,y) u = 0\,,
\end{equation}
which arise in numerous quantum-mechanical problems and as  approximate models for the wave propagation in the parabolic equation method and its generalizations
\cite{bab-bul}.
For  calculating solutions of this equation in unbounded domains it is essential to introduce
boundary conditions at the boundaries of the computational domain which
model the free transmission of the waves through these boundaries. As these boundary conditions must minimize
the amplitudes of waves reflected from boundaries, they are called absorbing boundary conditions \cite{eng-maj}.
\par
There is a significant number of works where the problem of constructing of such boundary
conditions was considered. The main approaches in these works consist either in factorization
of the differential operator of the equation under consideration into pseudodifferential factors, each of which describes
the unidirectional wave propagation \cite{eng-maj,shib,kus} and use the differential approximations of these factors
for the formulation of the boundary conditions, or in formulation of the absorbing boundary conditions as
the matching condition with the free space solution outside of the computational domain
(for the Eq.~(\ref{t1}) see the paper \cite{bas-pop}.

\par
As is known, the approximate description of the unidirectional propagation of the waves
can be obtained by the generalized multiple-scale method \cite{nay}, which in particular cases gives the same
results as the WKB or ray method. In this approach the algebraic factorization of the Hamilton-Jacobi equation
replaces the factorization of the differential operator. In the simplest case this approach was reported in
\cite{tr}.

\section{Derivation of the boundary conditions}
To apply the multiple-scale method \cite{nay} to the Eq.~(\ref{t1}) we
introduce the slow variables $X=\epsilon x$, $Y=\epsilon y$,
the fast variable $\eta = (1/\epsilon)\theta(X,Y)$, where $\epsilon$ is a small parameter,
change the partial derivatives in Eq.~(\ref{t1}) for the prolonged ones by the rules
$$
\frac{\partial}{\partial x} \rightarrow \epsilon\frac{\partial}{\partial X} +
\theta_X \frac{\partial}{\partial \eta}\,,\qquad
\frac{\partial}{\partial y} \rightarrow \epsilon\frac{\partial}{\partial Y} +
\theta_Y \frac{\partial}{\partial \eta}\,,
$$
and substitute in the obtained equation the expansion
$$
u = u_0 + \epsilon u_1 + \ldots\,.
$$

\par
Equating coefficients of like powers of $\epsilon$, we obtain at $O(\epsilon^0)$
the representation
\begin{equation*}
u_0 = A_0(X,Y)\mathrm{e}^{\mathrm{i}\eta}+B_0(X,Y)\mathrm{e}^{-\mathrm{i}\eta}\,,
\end{equation*}
and the Hamilton-Jacobi equation for the phase function $\theta$
\begin{equation} \label{t2}
\theta_X +  \beta(\theta_Y)^2 - \nu = 0\,.
\end{equation}
Later we will consider the one-way part of this solution
\begin{equation*}
u_0 = A_0\mathrm{e}^{\mathrm{i}\eta} = A_0(X,Y)\exp({\frac{\mathrm{i}}{\epsilon}\,\theta(X,Y)})\,.
\end{equation*}
\par
The solvability condition for the $O(\epsilon)$-equation
\begin{equation*}
\begin{split}
\mathrm{i}u_{0X} + \mathrm{i}\theta_X u_{1\eta} +&\beta\left(\theta_Y u_{0\eta}\right)_Y +\\
\beta\theta_Y u_{0Y\eta} + &\beta\left(\theta_Y\right)^2 u_{1\eta\eta} +\nu u_1 = 0\,,
\end{split}
\end{equation*}
regarded as a differential equation for $u_1$ with respect to the variable $\eta$ is
\begin{equation} \label{t3}
A_{0X}+\beta \theta_Y A_{0Y} +\beta(\theta_Y A_0)_Y = 0\,.
\end{equation}
Adding Eq.~(\ref{t2}), multiplied by $(\mathrm{i}/\epsilon)A_0\exp((\mathrm{i}/\epsilon)\theta)$, to Eq.~(\ref{t3}),
multiplied by  $\exp((\mathrm{i}/\epsilon)\theta)$, we obtain
$$
u_{0X} + 2\beta\theta_Y u_{0Y} - \frac{\mathrm{i}}{\epsilon}\beta (\theta_Y)^2 u_0+ \beta \theta_{YY} u_0
- \frac{\mathrm{i}}{\epsilon}\nu u_0 = 0\,,
$$
or, in initial coordinates $(x,y)$ and introducing the wave number $k=\theta_Y$, which is $O(1)$ in the method used,
we get finally
\begin{equation} \label{t4}
u_{0x} + 2\beta k u_{0y} - \mathrm{i} \beta k^2 u_0 +\beta k_y u_0- \mathrm{i}\nu u_0 = 0\,.
\end{equation}

\par
The system of Eqs.~(\ref{t2}) and (\ref{t4}) describes the geometric optic approximation to the Eq.~(\ref{t4}),
where two types of waves exist: propagating in positive direction along the $y$-axis, when $k>0$, and in negative
direction when $k<0$ (the turning points are excluded from consideration in this paper).
So we can use Eq.~(\ref{t4}), replacing $u_0$ by $u$, as an approximate non-reflecting boundary condition
at the boundaries of the form $\{(x,y)|y=const\}$.

\par
Formally we set the boundary value problem for Eq.~(\ref{t1}) in the strip $\{x,y\,|\,\,x\ge0\,,\, a\le y \le b\}$
(the initial-boundary value problem in the domain $a\le y \le b$, if $x$ is considered as the evolution variable),
specifying the  following boundary conditions:
\begin{equation*}
u=u_I\quad \text{at}\quad x=0\,,
\end{equation*}
\begin{equation} \label{t6}
\begin{split}
u_x - 2\beta|k| u_y - \mathrm{i} \beta k^2 u -\beta |k|_y - \mathrm{i}\nu u = 0\quad\text{at}\quad y=a\,, \\
u_x + 2\beta|k| u_y - \mathrm{i} \beta k^2 u +\beta |k|_y - \mathrm{i}\nu u = 0\quad\text{at}\quad y=b\,,
\end{split}
\end{equation}
where $k=\theta_Y$ for $\theta$, which is a solution of the Cauchy problem for the Hamilton-Jacobi
equation  (\ref{t2}) with the initial condition
\begin{equation*} 
\theta(X,Y)=\theta_I(Y)\quad\text{at}\quad X=0\,,
\end{equation*}
specified for all values of $Y$. This initial condition appears as a result of the representation
of initial data $u_I$ in the rapidly oscillating form
\begin{equation*} 
u_I=A_I(\epsilon y)\exp((\mathrm{i}/\epsilon)\theta_I(\epsilon y))\,.
\end{equation*}
Such a representation is not unique  and depends on auxiliary information that is used
for recognizing the small parameter  $\epsilon$, splitting the initial
condition into the amplitude and rapidly oscillating parts and determining the way of
extrapolation of  $\theta_I$ to all values of $Y$ and $\nu$ outside the strip (this is needed
to set the Cauchy problem for Eq.~(\ref{t2}))
\par
Now we compare Eq.~(\ref{t4})  with equations obtained by the formal factorization of Eq.~(\ref{t1}):
\begin{equation} \label{t7}
\begin{split}
 \mathrm{i} u_x +  \beta u_{yy} + \nu u \approx&
 \quad\mathrm{i}\left(\sqrt{\frac{\partial}{\partial x} - \mathrm{i}\nu} +
\sqrt{\mathrm{i}\beta}\frac{\partial}{\partial y}\right)\times\\
&\left(\sqrt{\frac{\partial}{\partial x} - \mathrm{i}\nu} -
\sqrt{\mathrm{i}\beta}\frac{\partial}{\partial y}\right) u =0\,,
\end{split}
\end{equation}
which has an approximate character, because the operators $\sqrt{\mathrm{i}\beta}\,\partial/\partial y$ and
$\sqrt{\partial/\partial x - \mathrm{i}\nu}$ do not commute in general.
The Pad\'e approximant \cite{be-orsz}  $P^1_0$ of the square root
$\sqrt{\partial/\partial x - \mathrm{i}\nu}$ with respect to $\partial/\partial x$ is
$$
\sqrt{\frac{\partial}{\partial x} - \mathrm{i}\nu} =
\sqrt{-\mathrm{i}\nu}\left(1+\frac{\mathrm{i}}{2\nu}\frac{\partial}{\partial x}\right)\,,
$$
and using it in factors of Eq.~(\ref{t7}), we obtain the equations
\begin{equation} \label{t8}
u_x\mp 2\sqrt{\beta\nu}\, u_y - \mathrm{i}\nu u=0\,.
\end{equation}
If we put that the phase function does not depends on $X$, $\theta_X=0$, then from
the Hamilton-Jacobi equation we get
$$
\theta_Y = \pm\sqrt{\frac{\nu}{\beta}}\,,\quad \beta (\theta_Y)^2 + \nu = 2\nu\,,
$$
and after substitution of these expressions into Eq.~(\ref{t4}) we obtain Eq.~(\ref{t8}).
Note that if the potential $\nu$ is vanished at the boundaries, the boundary conditions (\ref{t8})
degenerate to the Dirichlet conditions.
\par
Because of that in Shibata's paper \cite{shib} was used another linear approximation of the square root,
namely the linear interpolation between two points chosen without sufficient physical justification.
\par
Nevertheless, we will see later that our Eq.~(\ref{t4}), having as an approximation  the same linear nature,
works quite well.
\par
Returning to the multiple-scale expansion, we consider the next step, which leads to the generalization
of the rational-linear approximation discussed in Kuska's paper \cite{kus}.
\par
At $O(\epsilon^2)$ we obtain
\begin{equation*} 
\begin{split}
\mathrm{i}u_{1X} + \mathrm{i}\theta_X u_{2\eta} +&\beta\left(\theta_Y u_{1\eta}\right)_Y +\\
\beta\theta_Y u_{1Y\eta} + &\beta\left(\theta_Y\right)^2 u_{2\eta\eta} +\nu u_2 +\beta u_{0YY} = 0\,,
\end{split}
\end{equation*}
the solvability condition for which is
\begin{equation} \label{r2}
\mathrm{i}u_{1X} + \beta\left(\theta_Y u_{1\eta}\right)_Y +
\beta\theta_Y u_{1Y\eta} + \beta u_{0YY} = 0\,.
\end{equation}
Putting, on the strength of the same argumentation as in the $u_0$-case,
$$
u_1 = A_1\mathrm{e}^{\mathrm{i}\eta}\,,
$$
 we get from Eq.~(\ref{r2})
\begin{equation} \label{r3}
A_{1X} + \beta\left(\theta_Y A_1\right)_Y +
\beta\theta_Y A_{1Y} -\mathrm{i}\beta A_{0YY} = 0\,.
\end{equation}
In the same manner as Eq.~(\ref{t4}) was derived, we obtain from this
\begin{equation} \label{r4}
\begin{split}
u_{1x} + 2\beta k u_{1y} - &\mathrm{i} \beta k^2 u_1 +\beta k_y u_1- \mathrm{i}\nu u_1 - \\
&\epsilon\mathrm{i}\beta A_{0YY}\exp({\frac{\mathrm{i}}{\epsilon}\,\theta(X,Y)}) = 0\,.
\end{split}
\end{equation}
We introduce now the approximation of the first order as
$$
\bar u = u_0 + u_1
$$
and obtain the equation for this quantity from
\begin{equation*} 
\begin{split}
LHS&(\text{Eq.~(\ref{t4})}) + \epsilon LHS(\text{Eq.~(\ref{r4})}) -\\
&\epsilon^3 \mathrm{i}\beta A_{1YY}\exp({\frac{\mathrm{i}}{\epsilon}\,\theta(X,Y)})+O(\epsilon^3)=0\,,
\end{split}
\end{equation*}
where $LHS$ means `left-hand side', as usual.
Expressing the terms $\mathrm{i}\beta A_{0YY}\exp({\frac{\mathrm{i}}{\epsilon}\,\theta(X,Y)})$ and
$\mathrm{i}\beta A_{1YY}\exp({\frac{\mathrm{i}}{\epsilon}\,\theta(X,Y)})$ from differentiated
with respect to $Y$ Eqs~(\ref{t3}) and (\ref{r3}), after some algebra  we obtain
\begin{equation*} 
\begin{split}
&\mathrm{i}(\nu + 3\beta k^2) \bar u_y - \bar u_{xy} + 3\mathrm{i}k\bar u_x +
k(\beta k^2 + 3\nu)\bar u+\\
&\mathrm{i}\nu_y\bar u -\beta k_{yy}\bar u - 3\mathrm{i}\beta k k_y\bar u -6\beta k_y\bar u_y + O(\epsilon^3)=0\,.
\end{split}
\end{equation*}
Then, the equations
\begin{equation} \label{r6}
\begin{split}
&\mathrm{i}(\nu + 3\beta k^2) u_y -  u_{xy} \mp 3\mathrm{i}|k| u_x
\mp |k| (\beta k^2 + 3\nu) u +\\
&\mathrm{i}\nu_y u \pm \beta |k|_{yy} u \pm 3\mathrm{i}\beta |k| |k|_y u  \pm 6\beta |k|_y u_y =0
\end{split}
\end{equation}
can be proposed as the corresponding non-reflecting boundary conditions,
where the signs '$-$' and '$+$' corresponds to $y=a$ and $y=b$ respectively.
\par
The rational-linear boundary conditions from Kuska's paper, written in our notations, read
\begin{equation}\label{r7}
\mathrm{i}(\nu + 3\beta k^2) u_y -  u_{xy} \mp 3\mathrm{i}|k| u_x \mp |k| (\beta k^2 + 3\nu) u =0\,.
\end{equation}
These conditions were derived there by the factorization method with the
Pad\'e approximation \cite{be-orsz}  $P^1_1$ of the square root
$\sqrt{\partial/\partial x - \mathrm{i}\nu}$.
The last three terms of Eq.~(\ref{r6}) is absent in Eq.~(\ref{r7}) because $k$ is a constant in that paper.
The term $\mathrm{i}\nu_y u$ is absent due to the approximate character of the factorization method,
mentioned above.

\section{Numerical experiments}
As examples of application of the boundary conditions (\ref{t6}) and (\ref{r6}) we will
present the numerical simulation of the Gaussian beam
\begin{equation} \label{t9}
u(x,y) = \sqrt{\frac{x_0}{x+x_0}}\exp\left(\mathrm{i}\frac{y^2-yx_0(2y+px)}{4(x+x_0)} \right)\,,
\end{equation}
which was used also in the works \cite{kus,bas-pop}. As it is easily seen, the function from Eq.~ (\ref{t9})
is an exact solution of Eq.~(\ref{t1}) with  $\beta=1$ and $\nu=0$.
\par
Choose $x_0=-\mathrm{i}a$, where $a$ is a real number greater than zero. For the initial condition we have the expression
\begin{equation*} 
u_0 = u(0,y) = \exp\left(-\frac{y^2}{4a}\right)\exp\left(-\mathrm{i}\frac{py}{2}\right)\,,
\end{equation*}
and we choose as the initial phase
\begin{equation} \label{t11}
\theta_0 = \theta(0,y) = -\frac{1}{2}py\,.
\end{equation}
The solution of the Cauchy problem for the Hamilton-Jacobi equation (\ref{t2}) with the initial condition
(\ref{t11}) is (see, e.g. \cite{mas})
\begin{equation*} 
\theta(x,y) = \min_\xi{\left(\frac{(y-\xi)^2}{4x}-\frac{1}{2}p\xi\right)} =
 -\frac{1}{2}py-\frac{1}{4}p^2x\,,
\end{equation*}
and this solution we use in the boundary conditions (\ref{t6}) and (\ref{r6}). Note also that in this case
our first-order boundary conditions reduce to the Kuska's boundary conditions Eq.~(\ref{r7}).
The formal expansion of Eq.~(\ref{t9}) in powers of $1/\sqrt{a}$ shows that in this case the latter can be considered
 as a small parameter, which is confirmed by the results of calculations.

\begin{figure}
\includegraphics[width=0.45\textwidth]{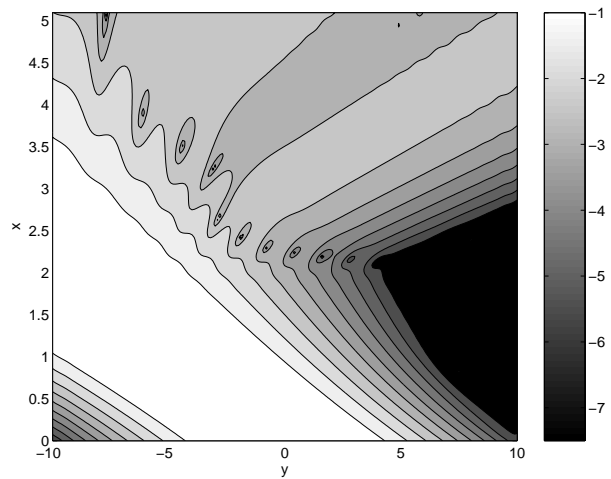}
\caption{Contour of the $\log_{10}(|u|)$ for the boundary conditions (\ref{t6}). $a=2$, $p=5$, $E/E_0=1.216\cdot10^{-4}$.
\label{fig1}}
\end{figure}

\begin{figure}
\includegraphics[width=0.45\textwidth]{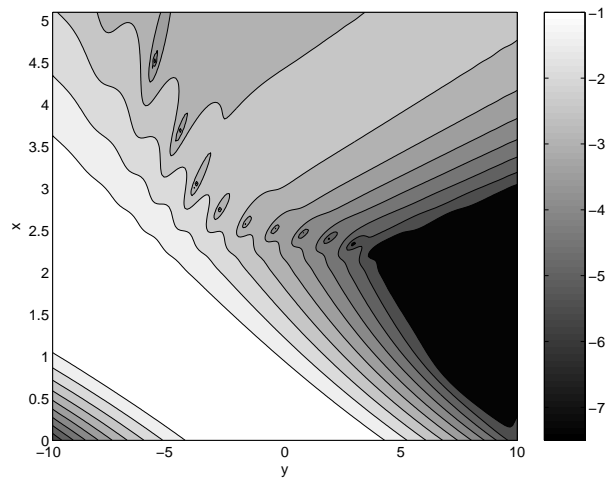}
\caption{Contour of the $\log_{10}(|u|)$ for the boundary conditions (\ref{r6}). $a=2$, $p=5$, $E/E_0=7.875\cdot10^{-5}$.
\label{fig2}}
\end{figure}

\begin{figure}
\includegraphics[width=0.45\textwidth]{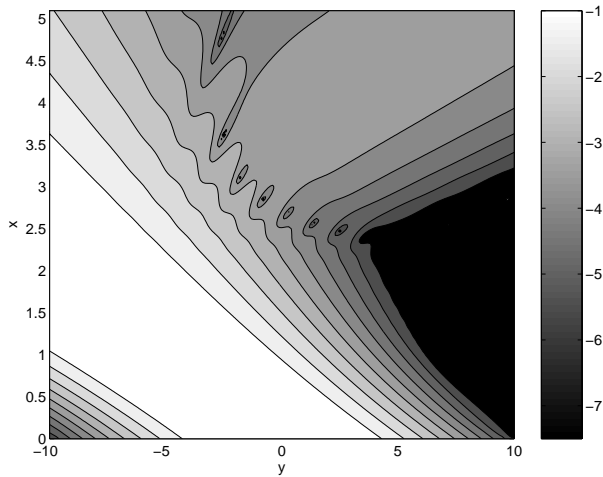}
\caption{Contour of the $\log_{10}(|u|)$  for the Baskakov-Popov boundary conditions \cite{bas-pop}.
 $a=2$, $p=5$, $E/E_0=3.251\cdot10^{-5}$.
\label{fig3}}
\end{figure}

\par
The calculations were done with the use of the Crank-Nicolson finite-difference scheme
 \cite{pot} with the parameters of the Gaussian beam  $a=2$, $p=5$ and $a=1/16$, $p=40$.
The results of calculations for the first case with the grid size $1025\times1025$  are presented
in FIG.~\ref{fig1} and  FIG.~\ref{fig2}.  In FIG.~\ref{fig3}
are presented the results of calculations with the use of the boundary conditions of Baskakov and Popov
\cite{bas-pop}, which are analytically exact, so this figure shows the effect of the used discretization.
In the captions of these figures we present also the values of the relative energy of the reflected waves
$$
E/E_0=\int^{y_{max}}_{y_{min}} |u(x_{max},y)|^2\,dy/\int^{y_{max}}_{y_{min}} |u(0,y)|^2\,dy\,.
$$

\par
The results of calculations for the narrow Gaussian beam with parameters $a=1/16$, $p=40$, which were done in the domain
$[0\le x\le 0.15]\times [-2\le y\le 2]$, show the following values of the relative energy:
\begin{itemize}
\item $3.585\cdot 10^{-5}$, $5.677\cdot 10^{-5}$ and $1.834\cdot 10^{-6}$ for the zeroth order,
first order and Baskakov-Popov
boundary conditions respectively on the grid $1025\times 1025$;
\item $1.066\cdot 10^{-4}$, $2.427\cdot 10^{-4}$ and $1.971\cdot 10^{-5}$ for the zeroth order,
first order and Baskakov-Popov
boundary conditions respectively on the grid $513\times 513$.
\end{itemize}
These results confirm that the inverse width of the beam $1/\sqrt{a}$ plays the r\^ole of the small parameter
and show also that for narrow beams the zeroth order boundary conditions can be better than the
first order ones, even on the big grid, and more robust with respect to the roughness
of the grid.
\par
The dependence of the accuracy of the boundary  conditions (\ref{r7}) on the beam width was investigated in some details
in \cite{kus}.

\section{Conclusion}
In this paper the absorbing boundary conditions (\ref{t6}) and (\ref{r6}) for the variable coefficient Schr\"odinger
type equation (\ref{t1}) were derived by the multiple-scale method. These boundary conditions explicitly take into account
the variability of the coefficients and are easy to use. The solution of the Hamilton-Jacobi equation, which is required
in these conditions, can be in most cases obtained analytically by the far-field approximations or numerically
by the method of the eulerian geometric optics \cite{be}.
\par
The reported method can be easily generalized to the many-dimensional case.

\begin{acknowledgments}
This work is supported by the Program No. 14 (part 2) of the Presidium of
the Russian Academy of Science.
\end{acknowledgments}

\end{document}